# Genome-wide organization of eukaryotic pre-initiation complex is influenced by nonconsensus protein-DNA binding

Ariel Afek and David B. Lukatsky*
*Department of Chemistry, Ben-Gurion University of the Negev, Beer-Sheva 84015, Israel*

## Abstract

Genome-wide binding preferences of the key components of eukaryotic pre-initiation complex (PIC) have been recently measured with high resolution in *Saccharomyces cerevisiae* by Rhee and Pugh (Nature (2012) **483**:295-301). Yet the rules determining the PIC binding specificity remain poorly understood. In this study we show that nonconsensus protein-DNA binding significantly influences PIC binding preferences. We estimate that such nonconsensus binding contribute statistically at least 2-3 kcal/mol (on average) of additional attractive free energy per protein, per core promoter region. The predicted attractive effect is particularly strong at repeated poly(dA:dT) and poly(dC:dG) tracts. Overall, the computed free energy landscape of nonconsensus protein-DNA binding shows strong correlation with the measured genome-wide PIC occupancy. Remarkably, statistical PIC binding preferences to both TFIID-dominated and SAGA-dominated genes correlate with the nonconsensus free energy landscape, yet these two groups of genes are distinguishable based on the average free energy profiles. We suggest that the predicted nonconsensus binding mechanism provides a genome-wide background for specific promoter elements, such as transcription factor binding sites, TATA-like elements, and specific binding of the PIC components to nucleosomes. We also show that nonconsensus binding influences transcriptional frequency genome-wide.

*Corresponding author
Email: lukatsky@bgu.ac.il

## Introduction

The assembly of the eukaryotic pre-initiation complex (PIC) is a critical step in the initiation of the transcription of eukaryotic genes (1-5). The PIC constitutes a multi-subunit protein complex; it assembles in the promoter regions of genes, in the vicinity of the transcription start site (TSS), and it regulates the transcription initiation by the RNA polymerase II enzyme (Pol II). The PIC consists of the TATA-binding protein (TBP), TBP-associated factors (TAFs), and general transcription factors (GTFs) (1-6). In a recent, seminal study, Rhee and Pugh measured for the first time, with unprecedented resolution, the genome-wide binding preferences of key components of the yeast PIC, such as TBP (Spt15), TFIIA (Toa2), TFIIB (Sua7), TFIID (Taf1), TFIIE (Tfa2), TFIIF (Tfg1), TFIIH (Ssl2), TFIIK (Kin28), and Pol II (Rpo21) (4). These measurements provide a remarkable snapshot of the *cis*-regulatory code of a eukaryotic genome.

The key and still open question is what rules do determine the DNA binding specificity of the PIC components? The answer to this question is impaired by the fact that for the majority of the GTFs in yeast, no consensus DNA binding sequence motifs have been identified (2, 4). One unambiguously identified core promoter element in yeast, with a relatively high specificity to the TBP, is the TATA box (7, 8). The initiator (INR) core promoter element has been also detected in yeast (9). In higher eukaryotes, yet not in yeast, additional core promoter elements have been identified (9-12).

Approximately 20% of yeast genes contain the TATA box motif (7, 8). These TATA-containing genes are highly regulated, they are associated with a response to stress, and predominantly utilize the SAGA complex (7, 8). However, in an apparent paradox, it was confirmed in (4) that TBP extensively binds the so-called TATA-less promoters, constituting approximately 80% of yeast genes and regulated by the TFIID complex (7, 8). The analysis has shown that the vast majority of such TATA-less promoters contain degenerate TATA-like elements (4). The latter finding highlights another key, long-standing question: what promoter sequence features do distinguish between the TATA-containing and TATA-less genes, or alternatively, between the SAGA-dominated and TFIID-dominated genes (6-8, 13)? It is well established that TFIID and SAGA complexes share several TAF subunits (8, 14). Therefore, yet another question is what promoter sequence features regulate the interplay between the specificity and redundancy (promiscuity) of TFIID and SAGA components?

We have recently suggested the existence of statistical, nonconsensus protein-DNA binding mechanism operating in eukaryotic genomes (15-17). We use the term 'nonconsensus protein-DNA binding' in order to describe the fact that the predicted protein-DNA binding free energy is computed without experimental knowledge of the high-affinity motifs for DNA-binding proteins. The predicted mechanism represents an extension of the notion of nonspecific protein-DNA binding introduced in seminal works of von Hippel, Berg et al. (18-22). In those works, nonspecific protein-DNA binding was schematically classified into two related mechanisms. The first mechanism is predominantly DNA sequence-independent and it assumes that DNA-binding proteins experience electrostatic attraction towards DNA, which is influenced by the overall DNA geometry (20). The second mechanism assumes that if a DNA-binding protein specifically binds to particular sequence motifs, then DNA sequences, which are similar to such specific motifs, will possess some enhanced protein-DNA binding affinity (20). We predicted that in addition to these two modes of nonspecific binding, there exists an enhanced statistical attraction between DNA-binding proteins and DNA sequences possessing particular symmetries and length-scales of sequence repeats. We use the term 'sequence correlations' in order to describe such sequence repeats (15). In particular, we showed that repeated homo-oligonucleotide tracts, such as poly(dA:dT) and poly(dC:dG), possess the strongest nonconsensus binding affinity towards DNA-binding proteins (15). In yeast, the computed genome-wide landscape of nonconsensus protein-DNA binding free energy significantly correlates with the experimentally measured nucleosome occupancy (16), and with statistical DNA binding preferences of approximately 200 transcription regulators (17).

In this study we seek to answer the question: how nonconsensus protein-DNA binding influences the binding preferences of the PIC? This paper is organized as follows. First, we define the precise notion of the free energy of nonconsensus protein-DNA binding, and estimate the genome-wide statistical strength of the effect, **Figure 1**. Second, we show that the genome-wide PIC occupancy is strongly correlated with the nonconsensus free energy landscape, **Figure 2**, **Figure 3**, and **Figure 4**. Third, we demonstrate that the PIC occupancy at both TFIID-dominated and SAGA-dominated groups of genes is in statistical agreement with the landscape of nonconsensus binding free energy, **Figure 5**. Yet, these two functionally different groups of genes can be distinguished based on their average free energy profiles. Fourth, we show that genome-wide transcriptional frequency is also influenced by nonconsensus protein-DNA binding, **Figure 6**. Finally, we notice that a significant fraction of yeast promoters possesses a secondary peak of the PIC occupancy, located in the upstream promoter region. The existence of this peak appears to be influenced by the enhanced occupancy of the -1 nucleosome, **Figure 7**.

# Results

### Definition of the free energy of nonconsensus protein-DNA binding

In order to compute the free energy of nonconsensus protein-DNA binding genome-wide in yeast, we first introduce a simple biophysical model of protein-DNA interactions (15). This model uses an ensemble of random DNA binders as a proxy for the phenomenon of nonspecific, nonconsensus protein-DNA binding in a crowded nuclear environment of a cell. By using the term 'random binder' we mean to express the fact that we do not use any experimentally predetermined protein-DNA binding preferences in order to model protein-DNA binding. The actual sequence of the yeast genome constitutes the only experimental, input parameter for our model. In particular, we assume that a protein makes contacts with $M$ DNA basepairs, and the protein-DNA interaction energy at each genomic position $i$:

$$U(i) = -\sum_{j=i}^{M+i-1} \sum_{\alpha=\{\mathrm{A,T,C,G}\}} K_\alpha s_\alpha(j), \qquad \text{Eq. (1)}$$

where for each genomic position $j$, $s_\alpha(j)$ represents the elements of a four-component vector of the type $(\delta_{\alpha A}, \delta_{\alpha T}, \delta_{\alpha C}, \delta_{\alpha G})$, where $\delta_{\alpha\beta} = 1$ if $\alpha = \beta$, or $\delta_{\alpha\beta} = 0$ if $\alpha \neq \beta$. For example, if the T nucleotide is positioned at the coordinate $j$ along the DNA, then this vector takes the form: $(0,1,0,0)$. The binding energy scale is set for each protein by the four parameters $K_\alpha$. In order to generate each model protein, we draw the values of $K_A$, $K_T$, $K_C$, and $K_G$ from the Gaussian probability distributions, $P(K_\alpha)$, with the zero mean, $\langle K_\alpha \rangle = 0$, and the standard deviation, $\sigma_\alpha = 2k_BT$, where $T$ is the temperature and $k_B$ is the Boltzmann constant. We have shown analytically in the past that the resulting free energy is qualitatively robust with respect to the choice of model parameters (15). This energy scale, $2k_BT \simeq 1.2$ kcal/mol, is chosen to represent a typical average strength of one hydrogen or electrostatic bond that a protein makes with a DNA bp in a cell (18, 20).

For each model protein, we define the partition function of protein-DNA biding within a sliding window of width $L = 50$ bp along the yeast genome:

$$Z = \sum_{i=1}^{L} \exp\left(-U(i) / k_B T\right), \qquad \text{Eq. (2)}$$

and the corresponding free energy of protein-DNA binding in this sliding window:

$$F = -k_B T \ln Z. \qquad \text{Eq. (3)}$$

We then assign the computed $F$ to the sequence coordinate in the middle of the sliding window. For example, for the chosen sliding window size, $L = 50\,\text{bp}$, 50 protein-DNA binding events contribute to the partition function, Eq. (2), in each sliding window, for each random binder. We verified that the resulting free energy landscape is qualitatively robust with respect to the choice of $L$ within a wide range of values, **Figure S1**. Moving the sliding window along the genome and computing $F$ at each genomic location, allows us to assign the free energy of nonconsensus protein-DNA binding to each DNA bp within the entire genome.

Next, we repeat the described procedure for an ensemble of 250 model random binders, and compute the average free energy, $\langle F \rangle_{\text{TF}}$, with respect to this ensemble, in each genomic location. The resulting free energy landscape, $\langle F \rangle_{\text{TF}}$, represents the statistical propensity of genomic DNA towards nonconsensus protein-DNA binding. The lower $\langle F \rangle_{\text{TF}}$ in a given genomic location, the stronger the attraction that DNA-binding proteins experience (on average) towards this location. We have shown previously that the predicted effect is entropy dominated, and it is driven by the correlation properties of the DNA sequence, rather than by the average sequence composition (15). In particular, genomic regions enriched in repeated poly(dA:dT) or poly(dC:dG) tracts, possess the strongest propensity (the lowest $\langle F \rangle_{\text{TF}}$) towards nonconsensus protein-DNA binding (15-17). This general, statistical effect stems from the symmetry of DNA, and intuitively, it can be understood in the following way. The dominant, attractive contribution to the partition function *Z*, Eq. (2), comes from the low-energy tail of the probability distribution for the protein-DNA interaction energies, *P*(*U*), Eq. (1). A protein moving along the DNA enriched in repeated poly(dA:dT) and/or poly(dC:dG) tracts, will possess a statistically wider distribution, *P*(*U*), compared with the case when the DNA sequence is either random, or has a different symmetry, such as, for example, TATATATAT..., or similar sequences. Such a wider distribution will statistically result in a lower free energy, Eq. (3). This effect is entropic, since it depends on the variation (fluctuation) of *U*, and not on the average value, $<U>$ (23). The latter property is also the reason for the fact that the free energy profiles are statistically robust with respect to the global variation of the nucleotide composition along the yeast genome, **Figure S1**.

In order to estimate the strength of the effect, we compute the probability distribution of the free energy difference in the vicinity of the TSSs, $\langle \Delta F \rangle_{\text{TF}} = \langle F_{\min} \rangle_{\text{TF}} - \langle F_{\max} \rangle_{\text{TF}}$, for 6,045 transcripts from (4), **Figure 1**. The position of the peak of this distribution, gives the average strength of the effect: $\langle \Delta F \rangle_{\text{TF}} \simeq -4.3\; k_B T \simeq -2.6\,\text{kcal/mol}$, per protein, per gene, on average, assuming that each protein makes contacts with *M*=8 bp upon DNA sliding. The resulting free energy profiles are statistically robust with respect to a moderate variation of the value of *M*, within a typical range of the TF binding site size in yeast, **Figure S1**. For the vast majority of genes, the minimum, $\langle F_{\min} \rangle_{\text{TF}}$, is located within the interval (-150,0) around the TSS. Intuitively, the estimated value means that DNA-binding proteins are statistically attracted towards the location of the free energy minimum within the promoter, and each protein gains statistically (on average) approximately $-3$ kcal/mol, exclusively due to the existence of nonconsensus protein-DNA binding.

## Nonconsensus protein-DNA binding influences genomic organization of the PIC

We now set out to seek an answer to the key question: how does the predicted nonconsensus protein-DNA binding affect the experimentally measured binding preferences of the PIC components in yeast genome-wide (4)? In order to answer this question, we compare the experimentally measured PIC occupancy in the vicinity of the TSSs for $\sim 4{,}000$ yeast genes (4), with the computed free energy landscape, **Figure 2 A** and **B**. The obtained statistically significant correlation suggests that nonconsensus protein-DNA binding significantly influences the PIC occupancy profile genome-wide, **Figure 2 A** and **B,** and **Figure S2**. The strongest effect is observed in the upstream promoter regions, in the immediate vicinity of the TSSs for the majority of genes. The lower the free energy, the stronger the statistical attraction towards DNA, that proteins experience due to nonconsensus protein-DNA binding. It is remarkable that the peak of the average PIC occupancy is shifted $\sim 50\,\text{bp}$ downstream relative to the average free energy minimum, **Figure 2 A**. This result is robust with respect to the choice of the sliding window size, $L$, within a wide range of values, **Figure S1**. The reason for this shift appears to be the interplay between the predicted nonconsensus binding, and specific, cooperative binding of the PIC complex to TATA-like elements in the core promoter regions (see below).

The genome-wide nucleosome occupancy profile determined in (4) shows statistically strong, positive correlation with the computed free energy landscape, **Figure 2 C** and **D**. The nucleosome occupancy is dramatically reduced exactly in the region of the reduced free energy, **Figure 2 C**. This is in agreement with our previous work (16), which used a different experimental source of nucleosome occupancy (24). Briefly, the proposed mechanism influencing nucleosome depletion stems from the competition between nucleosomes and transcription factors for binding to genomic regions characterized by the reduced free energy of nonconsensus protein-DNA binding (16). The effect of the PIC occupancy enrichment and the nucleosome occupancy depletion in the regions of the reduced free energy is clearly observable at the single-gene level, **Figure 3 A** and **B**. Notably, around 1,000 inverted genes exhibit a double-well free energy landscape in agreement with the corresponding PIC occupancy and nucleosome occupancy profiles, **Figure 3 B**.

The individual, average occupancy profiles of eight out of nine proteins analyzed in (4) are significantly, negatively correlated with the free energy landscape in a wide region around the TSSs, **Figure 4**. Intuitively, such negative correlation means that individual GTFs are attracted towards genomic regions possessing the reduced free energy. Interestingly, the Pol II occupancy is positively correlated with the free energy, **Figure 4**, alike the nucleosome occupancy. This observation can be possibly rationalized by the fact that Pol II may interact with +1 nucleosomes, and it should therefore resemble the nucleosome occupancy profile (4), which is positively correlated with the free energy, **Figure 2 D**. In addition, Pol II is recruited to the core promoters indirectly, through its specific interaction with GTFs, and thus specific binding dominates its occupancy in the immediate vicinity of the TSSs.

It is remarkable that among all GTFs, the average TFIID occupancy exhibits the weakest correlation with the free energy, **Figure 4**. This can be explained by the fact that TFIID experiences two competing interactions. It interacts attractively with the +1 nucleosome located downstream of it (4), and at the same time, TFIID is attracted towards the free energy minimum located upstream of it, **Figure 4**. We conclude therefore that the predicted nonconsensus protein-DNA binding free energy landscape significantly influences binding preferences of GTFs in promoter regions, genome-wide in yeast. This nonconsensus binding mechanism provides a background for specific promoter elements, such as transcription factor binding sites (TFBSs), TATA-like elements, INR elements, and specific binding of GTFs (such as TFIID) to nucleosomes.

We stress that contrary to the case of specific protein-DNA binding, nonconsensus binding is operational globally, within wide genomic regions. In particular, we verified that the predicted nonconsensus protein-DNA binding influences the GTF occupancy and the nucleosome occupancy around the 3' open reading frame (ORF) ends. **Figure S3** shows that the free energy is statistically significantly, negatively correlated with the PIC occupancy and positively correlated with the nucleosome occupancy, similar to the trends observed around the TSSs, **Figure 2**. Remarkably, at the single-gene level, the GTF occupancy profile of 1,860 tandem mRNA genes follows the free energy profile, **Figure S3**. Strikingly, even the individual occupancies of GTFs measured in (4) are significantly correlated with the free energy, **Figure S4**. This analysis leads us to conclude that nonconsensus protein-DNA binding mechanism operates and influences the genome-wide GTF occupancy and nucleosome occupancy within the wide genomic range, not only around the TSSs, but also around the ORF gene ends.

## Free energy landscape of nonconsensus protein-DNA interactions distinguishes between TFIID-enriched and TFIID-depleted promoters

Transcriptional regulation in yeast appears to be mechanistically bipolar: nearly 90% of the yeast genes are regulated by the TFIID complex, while the rest ~10% are regulated by the SAGA complex (8). The majority of TFIID-dominated genes are classified as TATA-less and housekeeping, while the majority of SAGA-dominated genes are TATA-containing and stress response (4, 7, 8). Remarkably, there exists a considerable crosstalk (redundancy) between the components of TFIID and SAGA complexes (8, 25).

With this in mind, we now ask the question how does the predicted nonconsensus protein-DNA binding free energy landscape affect the PIC occupancy in the TFIID-dominated and SAGA-dominated genes, respectively? **Figure 5 A** and **C** show that the PIC occupancy within both groups of genes measured in (4) is negatively correlated with the free energy. The nucleosome occupancy in these two groups of genes is positively correlated with the free energy, consistent with our previous analysis (16), **Figure 5 B** and **D**. While both groups of genes are affected by the nonconsensus binding, yet the average free energy profiles are clearly distinguishable between the two groups, **Figure 5 E**, where the TFIID-dominated (Taf1-enriched) genes are characterized by the narrower free energy landscape than SAGA-dominated (Taf1-depleted genes). We verified that TATA-less and TATA-containing genes behave quantitatively similar to TFIID-dominated and SAGA-dominated genes, respectively, **Figure S5**. Based on these observations, we conclude that nonconsensus protein-DNA binding statistically influences the PIC occupancy within the vast majority of the yeast genome, including both TFIID-dominated and SAGA-dominated genes. Yet, these two groups of genes are characterized by distinguishable average free energy landscapes. These findings lead us to a remarkable conclusion that the observed crosstalk and functional redundancy between the components of TFIID and SAGA complexes (8, 25), might originate, at least partially, from nonconsensus protein-DNA binding, intrinsically encoded into genomic DNA. Such nonconsensus binding affects, statistically, both TFIID and SAGA complexes, **Figure 5 A** and **C**.

We stress that our simple biophysical model of protein-DNA interactions does not use any experimental knowledge of the high affinity (consensus) protein-DNA binding sites or TATA-like box preferences, and therefore the computed nonconsensus free energy does include any contribution of sequence-specific (consensus) protein-DNA binding. Taking such sequence-specific contribution into account might shed light on the question of a relative significance of specific (consensus) versus nonconsensus effects for binding preferences of TFIID and SAGA complexes, respectively.

### Nonconsensus protein-DNA binding influences transcriptional frequency

We now proceed to quantify how the predicted nonconsensus protein-DNA binding influences gene expression on the genome-wide scale. **Figure 6** shows a statistically significant correlation between the computed free energy and the measured transcriptional frequency of ~4,000 genes (6). Genes with the reduced free energy of nonconsensus protein-DNA binding in the promoter regions, and hence with the higher levels of the GTF occupancy, exhibit statistically higher levels of transcriptional frequency. The fact that the observed correlation is only moderately strong emphasizes the great significance of other factors influencing gene expression, and first of all, the effect of specific TF-DNA binding that we do not take into account in the presented model.

### Nucleosomes flanking upstream promoter region influence PIC occupancy

The analysis of experimentally measured PIC binding preferences shows that a significant fraction of yeast promoters possesses a secondary peak in the upstream promoter regions, **Figure 7**. It is remarkable that the two groups of genes (a group with only a single peak, and a group with two peaks) are characterized by statistically indistinguishable average free energy profiles, and TATA-like element occupancy profiles (Materials and Methods and **Figure S6**). The initiator (INR) promoter element occupancy profiles (9, 12) are also indistinguishable between these two groups of genes, **Figure S6**. However, the -1 nucleosome occupancy is significantly enhanced in the group possessing the second peak in the upstream promoter region, **Figure 7**. This observation suggests that specific binding of the PIC components (such as Taf1) to the -1 nucleosome is an important additional factor regulating the PIC occupancy profiles.

## Discussion and conclusion

Here we predicted that the yeast genomic DNA exerts the nonconsensus protein-DNA binding potential acting, statistically, on all DNA-binding proteins, and in particular, on the GTFs. We described the action of this effective potential by assigning the free energy of nonconsensus protein-DNA binding to each genomic location. We then observed that the experimentally measured binding preferences of GTFs (4) behave in a remarkable agreement with the predicted free energy landscape, **Figure 2** and **Figure 4**. We estimated that the strength of the effect is, at least, 2-3 kcal/mol per one protein (on average). This value represents an additional attractive protein-DNA binding free energy gained by a protein (on average) in a promoter region exclusively due to nonconsensus protein-DNA binding. The predicted attractive effect is particularly strong at repeated poly(dA:dT) and poly(dC:dG) tracts. We emphasize that our simple biophysical model of protein-DNA interactions does not use any experimental knowledge of the high affinity (consensus) protein-DNA binding sites, and therefore it does not have fitting parameters. Despite the simplicity of the model, we suggest that our conclusions are quite general, and most likely, the predicted mechanism influencing PIC binding preferences, is operational in other eukaryotic genomes.

We observed that TFIID-enriched and TFIID-depleted (SAGA-dominated) genes are statistically distinguishable based on the nonconsensus protein-DNA binding free energy landscapes of these two groups, **Figure 5**. However, remarkably, nonconsensus protein-DNA binding influences the PIC occupancy of both TFIID-dominated and SAGA-dominated genes. In particular, the experimentally measured occupancies of TFIID-enriched and TFIID-depleted genes both show a significant correlation with the free energy landscape, **Figure 5 A** and **C**. This suggests that the predicted nonconsensus protein-DNA binding might be responsible for the observed crosstalk and functional redundancy between the TFIID and SAGA complexes (8, 14). We also observed that transcriptional frequency is correlated with the predicted free energy landscape, **Figure 6**. The fact that the observed correlation is not strong, although highly significant, highlights the importance of the specific protein-DNA binding component in transcriptional regulation.

We note that in a recent experimental study performed in vitro (26), it was shown that in addition to conventional TATA box binding, the TBP extensively binds poly(T) stretches. This effect might be also a direct consequence of the nonconsensus protein-DNA binding mechanisms predicted here. Additional experiments with poly(dA:dT) and poly(dC:dG) stretches might provide a further insight into the mechanism of the observed effect.

In conclusion, the predicted nonconsensus protein-DNA binding constitutes a genome-wide attractive background (sink), globally modulating the statistical occupancy of transcription factors (and other DNA-binding proteins) along the genome. Contrary to specific protein-DNA binding, relatively long genomic regions (of at least few tens of bp) contribute to this effect. Despite such intrinsic non-locality, we observed that, statistically, nonconsensus binding significantly influences binding preferences of the majority of the PIC components in core promoter regions genome-wide. This suggests that intrinsically encoded nonconsensus protein-DNA binding might be tightly linked to specific protein-DNA binding in fine-tuning transcriptional regulation.

## Materials and Methods

### Occupancy of individual GTFs

The experimentally measured occupancies of individual GTFs determined by the ChIP-exo method are taken from (4).

### *p*-value calculations

In order to compute the *p*-value for **Figure 5**, we generated 10,000 pairs of randomly chosen groups of 4,755 and 1,135 genes, respectively, representing random replicas of TAF1-enriched and TAF1-depleted gene groups, respectively. For each of these pairs we calculated the difference between the average free energy of TAF1-enriched and TAF1-depleted groups within the range $(-400, 400)$ around the TSS. The probability that the computed value of the difference in the randomized sets is larger than the actual value is assigned as the *p*-value. The *p*-value for **Figure S5** is computed analogously. In order to compute the *p*-value for **Figure 7**, we first selected 10,000 pairs of randomly chosen groups of 1,432 and 2,513 genes, respectively, representing randomized analogs for the actual double-peak and the single-peak groups, respectively. Next, for each pair we calculated the absolute difference in the peak value of the average -1 nucleosome occupancy between the double-peak and the single-peak groups, respectively. Finally, we computed the probability that this difference reaches the actual value. This probability was then assigned as the *p*-value.

### TATA-containing and TATA-less genes

The definitions of TATA-containing and TATA-less genes, as well as TAF1-enriched and TAF1-depleted genes are adopted from (4).

### TATA-like box occupancy score

In order to assign the TATA-like box occupancy score in **Figure 7,** we used the definition from (4). In particular, we searched for the conventional TATA-like motif, TATA(A/T)A(A/T)(A/G). We assigned a score 8 for a perfect match to this motif, score 7 to a match with one mismatch, 6 to a match with two mismatches, and 0 otherwise. We verified that an alternative definition of the TATA-like box occupancy score, based on PWM (12), leads to similar conclusions, **Figure S6**.

## Acknowledgements

We thank Ho Sung Rhee for helping us with compiling the ChIP-exo data. D.B.L. acknowledges the financial support from the Israel Science Foundation (ISF) grant 1014/09. A.A. is a recipient of the Negev Faran graduate fellowship, and the Lewiner graduate fellowship.

## Supporting Material

Six supporting figures.

## Figure Legends

**Figure 1**. The nonconsensus protein-DNA binding free energy is statistically reduced in the yeast promoter regions. Computed probability distribution, $P(\Delta f)$, of the free energy difference per bp, $\Delta f = f_{\min} - f_{\max}$, for each transcript from (4), where $f_{\min}$ and $f_{\max}$ are the minimal and the maximal free energy values, respectively, in the interval $(-400,400)$ around the TSS, where we defined, $\Delta f = \langle \Delta F \rangle_{\mathrm{TF}} / M$, and we used $M$=8. $P(\Delta f)$ is computed based on 6,045 transcripts from (4). The average value, $\langle \Delta f \rangle = -0.54\ k_B T$. The inset shows an example of the computed free energy profile, $f = \langle F \rangle_{\mathrm{TF}} / M$, for the CDC15 gene, with the definitions of $f_{\min}$, $f_{\max}$, and $\Delta f$.

**Figure 2.** The free energy of nonconsensus TF-DNA binding negatively correlates with the combined GTF occupancy, and positively correlates with the nucleosome occupancy. (**A**) The average free energy of nonconsensus TF-DNA binding per bp, $\langle f \rangle = \left\langle \langle F \rangle_{\mathrm{TF}} \right\rangle_{seq} / M$, (blue), and the average, combined occupancy profile of all GTFs from (4) (red), around the TSSs of 3,945 genes. The notation $\langle \mathrm{PIC} \rangle$ describes the average, combined occupancy profile of all nine GTFs. The linear correlation coefficient is computed for a linear fit of $\langle f \rangle$ versus the average, combined GTF occupancy at individual genomic locations, every 20 bp, within the interval $(-990,990)$. In order to compute error bars, we divided genes into five randomly chosen subgroups, and computed $\langle f \rangle$ for each subgroup. The error bars are defined as one standard deviation of $\langle f \rangle$ between the subgroups. The error bars for the combined GTF occupancy are computed analogously. (**B**) Correlation between the minimal value of the free energy of nonconsensus TF-DNA binding, $f_{\min} = \min(f)$, and the combined occupancy of all GTFs, computed for individual genes in non-overlapping windows of 80 bp within the entire interval $(-990,990)$ around the TSS for each of these 3,945 genes. The data are binned into 50 bins. We verified that a correlation between the computed free energy profiles and the experimentally determined PIC occupancy remains statistically significant for a narrower range around the TSS, **Figure S2**. (**C**) The average free energy of nonconsensus TF-DNA binding per bp, $\langle f \rangle$ (blue), and the average nucleosome occupancy (4) (red), around the TSSs of 3,945 genes. (**D**) Correlation between the minimal value of the free energy of nonconsensus TF-DNA binding, $f_{\min} = \min(f)$, and the average nucleosome occupancy computed for individual genes, computed in non-overlapping windows of 80 bp within the entire interval $(-990,990)$ around the TSS for each of 3,945 genes. The data are binned into 50 bins.

**Figure 3.** The heat maps demonstrate that at the individual gene level, the free energy of nonconsensus TF-DNA binding negatively correlates with the combined GTF occupancy, and positively correlates with the nucleosome occupancy around the TSSs. (**A**) The heat maps represent the free energy of nonconsensus TF-DNA binding per bp, $f$, the combined occupancy of the GTFs, and the nucleosome occupancy, respectively, for individual genes aligned with respect to the TSS. (**B**) The heat maps represent the free energy of nonconsensus TF-DNA binding, $f$, the combined occupancy of the GTFs, and the nucleosome occupancy, respectively, for 1,078 inverted mRNA genes aligned with respect to the TSS. The genes are sorted by intergenic length.

**Figure 4.** The measured occupancy profiles of individual GTFs are significantly affected by nonconsensus TF-DNA binding. The average free energy of nonconsensus TF-DNA binding per bp, $\langle f\rangle = \langle\langle F\rangle_{\mathrm{TF}}\rangle_{seq} / M$ (blue), and the average occupancy profile of individual GTFs from (4) (red), around the TSSs of 3,945 genes. In order to compute error bars, we divided genes into five randomly chosen subgroups, and computed $\langle f\rangle$ for each subgroup. The error bars are defined as one standard deviation of $\langle f\rangle$ between the subgroups. The error bars for the GTF occupancy are computed analogously. We used $M = 8$ in all calculations. The linear correlation coefficient is computed in each case for a linear fit of $\langle f\rangle$ versus the average GTF occupancy at individual genomic locations, every 20 bp, within the interval $(-990,990)$.

**Figure 5.** Statistical PIC binding preferences to both TFIID-dominated and SAGA-dominated genes negatively correlate with the nonconsensus free energy landscape, yet these two groups of genes are distinguishable based on the average free energy profiles. (**A**) Correlation between the minimal value of the free energy of nonconsensus TF-DNA binding, $f_{\min} = \min(f)$, and the average GTF occupancy of 3,068 TAF1-enriched genes. The correlation is computed for individual genes in non-overlapping windows of 80 bp within the entire interval $(-990,990)$ around the TSS. The data are then binned into 50 bins. (**B**) Similar to (**A**), but now $f_{\min}$ is correlated with the nucleosome occupancy of these TAF1-enriched genes. (**C**) Correlation between $f_{\min}$, and the average GTF occupancy of 877 TAF1-depleted genes. The correlation is computed for individual genes in non-overlapping windows of 80 bp within the entire interval $(-990,990)$ around the TSS. The data are binned into 50 bins. (**D**) Similar to (**C**), but now $f_{\min}$ is correlated with the nucleosome occupancy of these TAF1-depleted genes. (**E**) The average free energy of nonconsensus TF-DNA binding per bp, $\langle f\rangle$, for a larger set of 4,755 TAF1-enriched genes (blue), and 1135 TAF1-depleted genes (red), around the TSSs.

**Figure 6.** Nonconsensus TF-DNA binding influences transcriptional frequency genome-wide. (**A**) Correlation between the minimal value of the free energy of nonconsensus TF-DNA binding, $f_{\min}$, and the transcriptional frequency from (6), computed for 3,811 genes. For each gene, $f_{\min}$ is computed in the interval $(-150,0)$ around the TSS. The data are binned into 45 bins, ordered by the magnitude of the transcriptional frequency. The outlier point corresponding to the highest frequency bin is shown in grey. Removing this point significantly improves the correlation coefficient. (**B**) Correlation between the experimentally measured in (4) peak occupancy of TFIIB in the promoter region and the transcriptional frequency for these 3,811 genes.

**Figure 7.** Nucleosomes flanking upstream promoter region influence PIC occupancy. Specific binding of the PIC components to the -1 nucleosome might be responsible for the emergence of a secondary peak in the PIC occupancy profiles. Left panel from top to bottom: The heat map represents the combined occupancy of the GTFs (we use the term "PIC occupancy" to describe the latter) in the genes selected with a condition of the existence of a second peak in the combined GTF occupancy per gene, as measured in (4). Only the genes with the absolute upstream peak occupancy larger than 40 (in the occupancy score units used in (4)), and with a value of at least 50%, as compared with the downstream peak occupancy were selected. As a result, 1,432 double-peak (left panel) and the rest of 2,513 single-peak (right panel) genes were selected. The next graphs (from top to bottom) represent the average combined occupancy of the GTFs, $< \mathrm{PIC} >$; the average free energy of nonconsensus TF-DNA binding

per bp, $\langle f \rangle$; the average TATA-like box occupancy score (Materials and Methods); and the average nucleosome occupancy, $\langle \mathrm{NO} \rangle$, respectively. Right panel shows analogous graphs for the rest of 2,513 single-peak genes. The bottom-right graph shows the absolute difference of the average, maximal values of the -1 nucleosome occupancy between the double-peak and the single-peak groups, respectively, as well as the computed *p*-value for this difference.

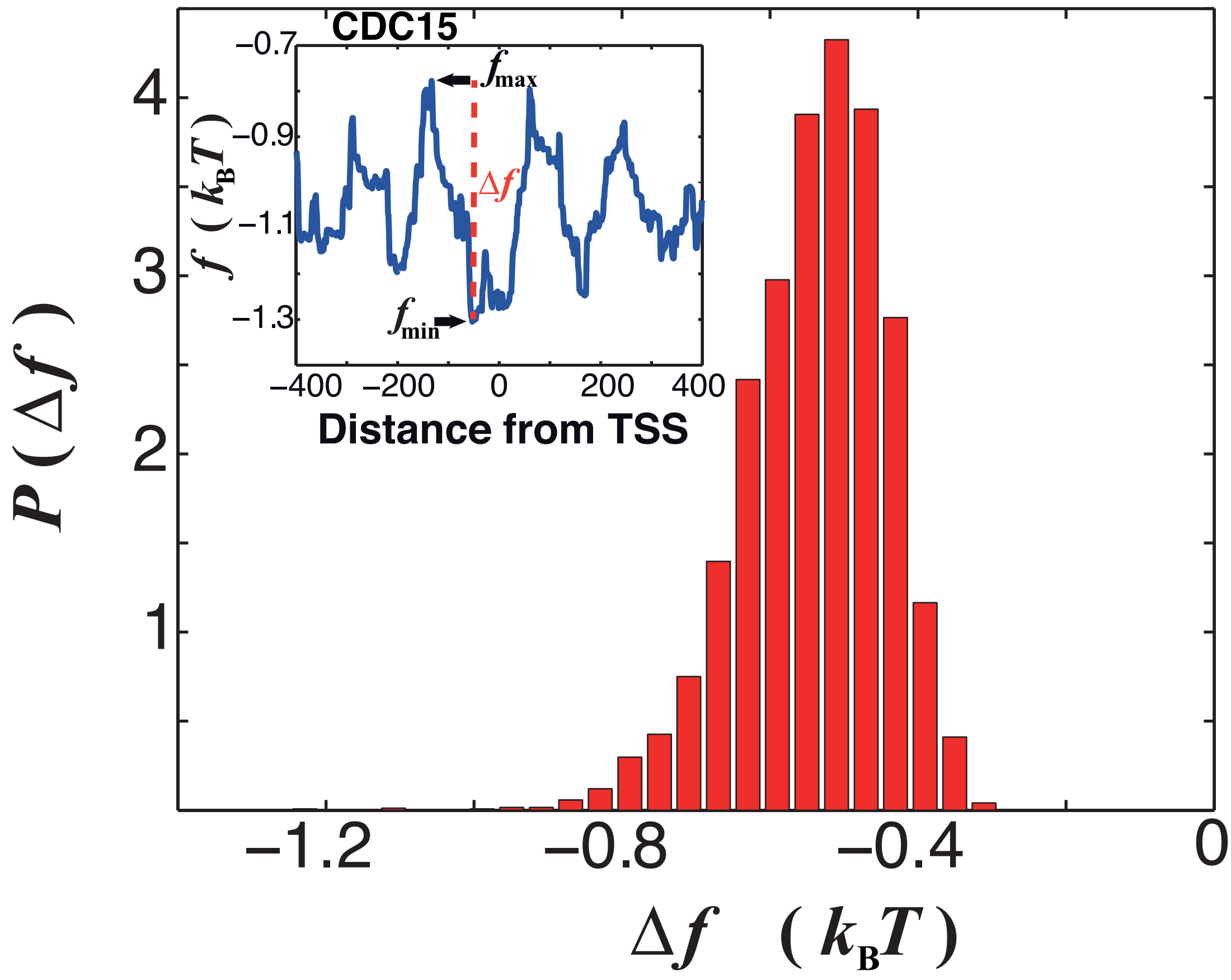

**Figure 1**

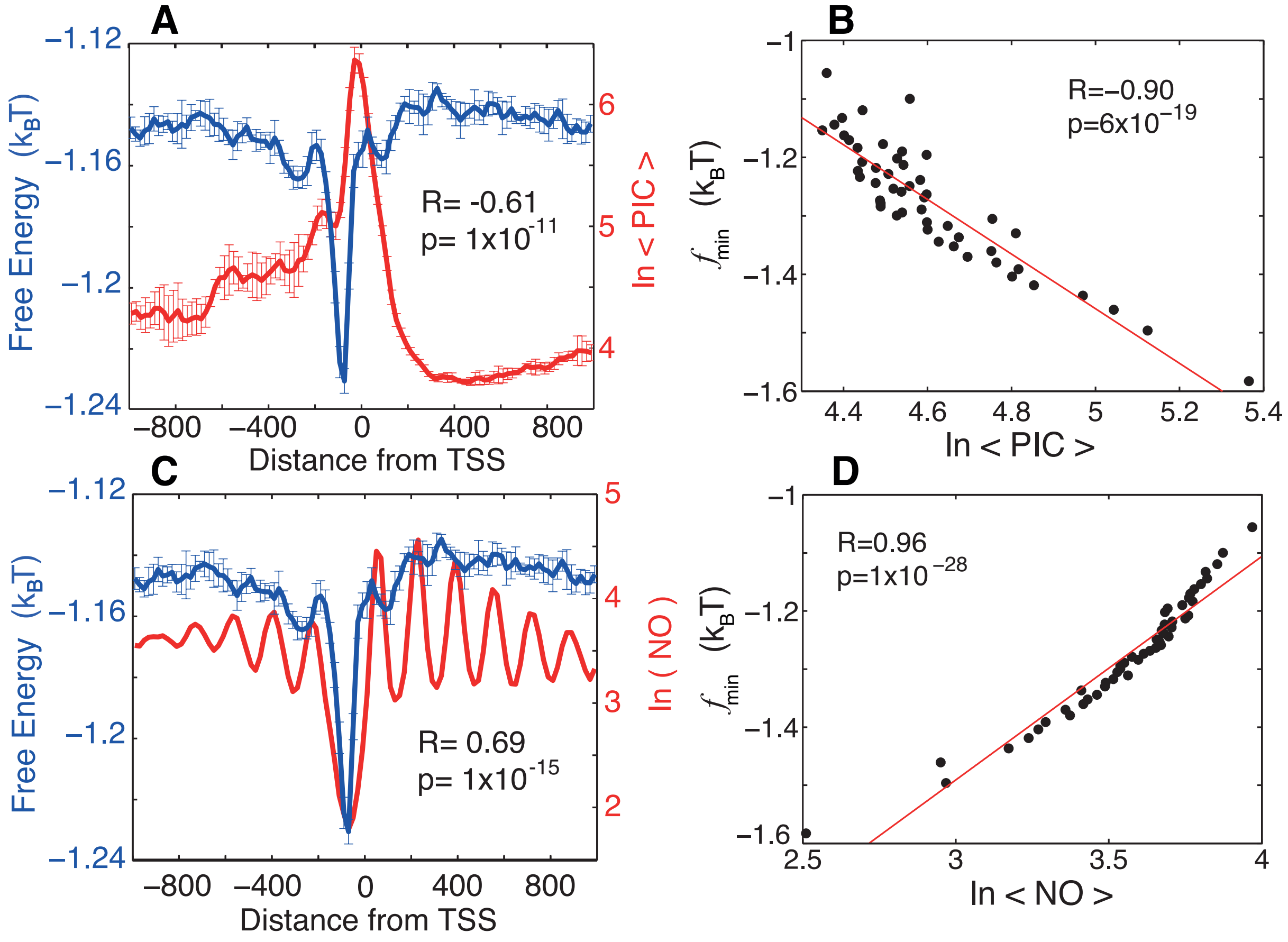

**Figure 2**

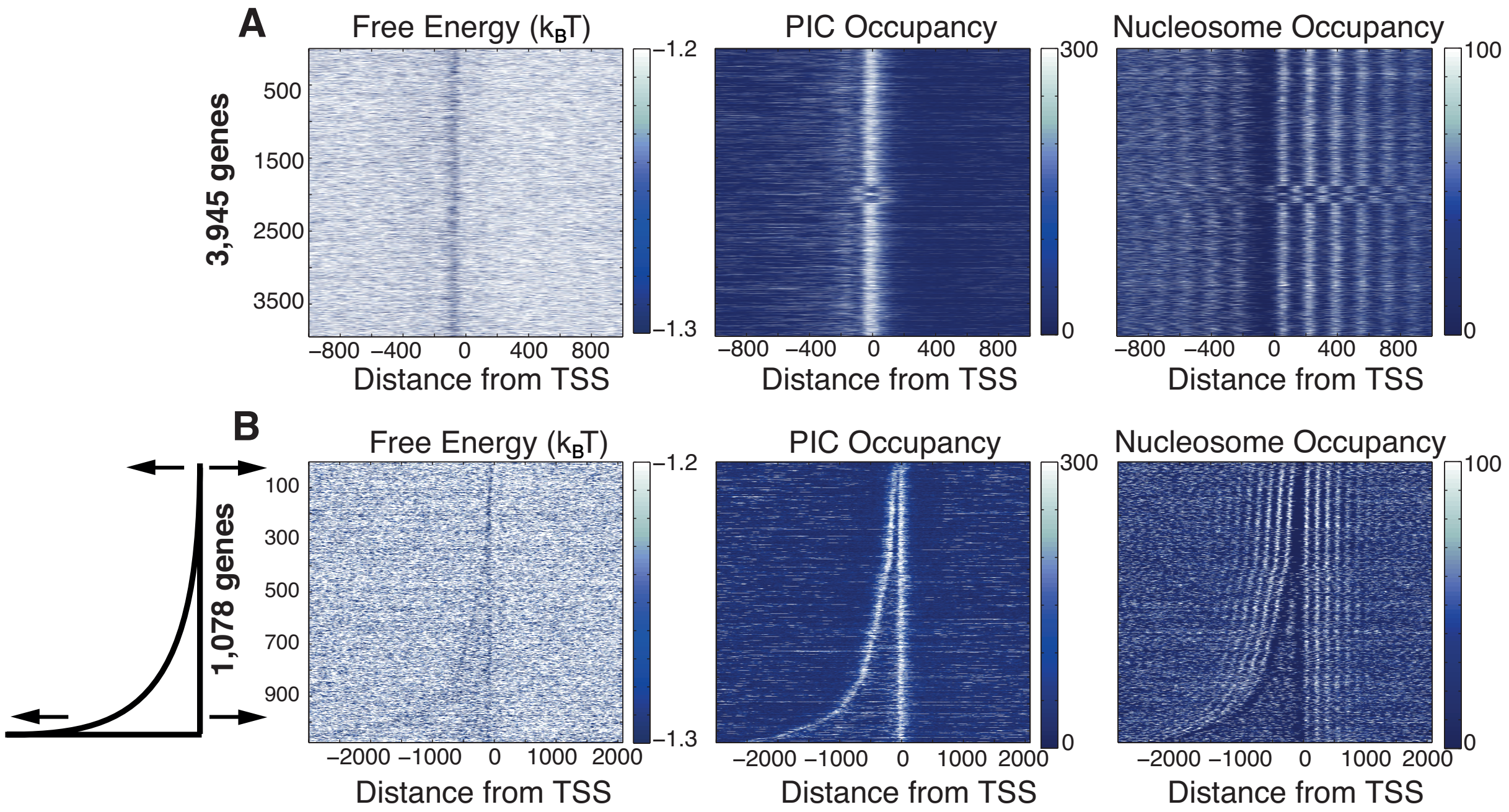

**Figure 3**

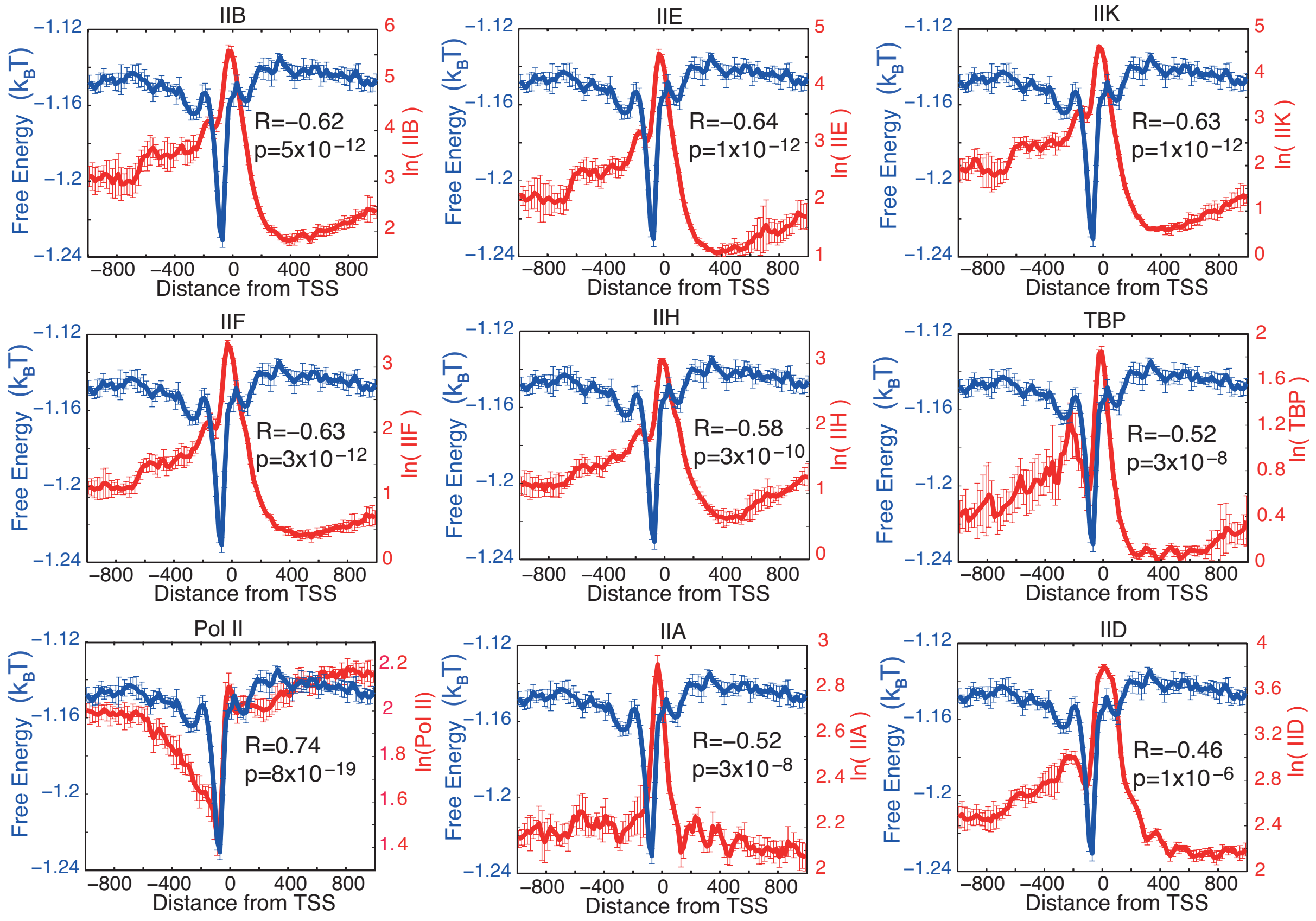

**Figure 4**

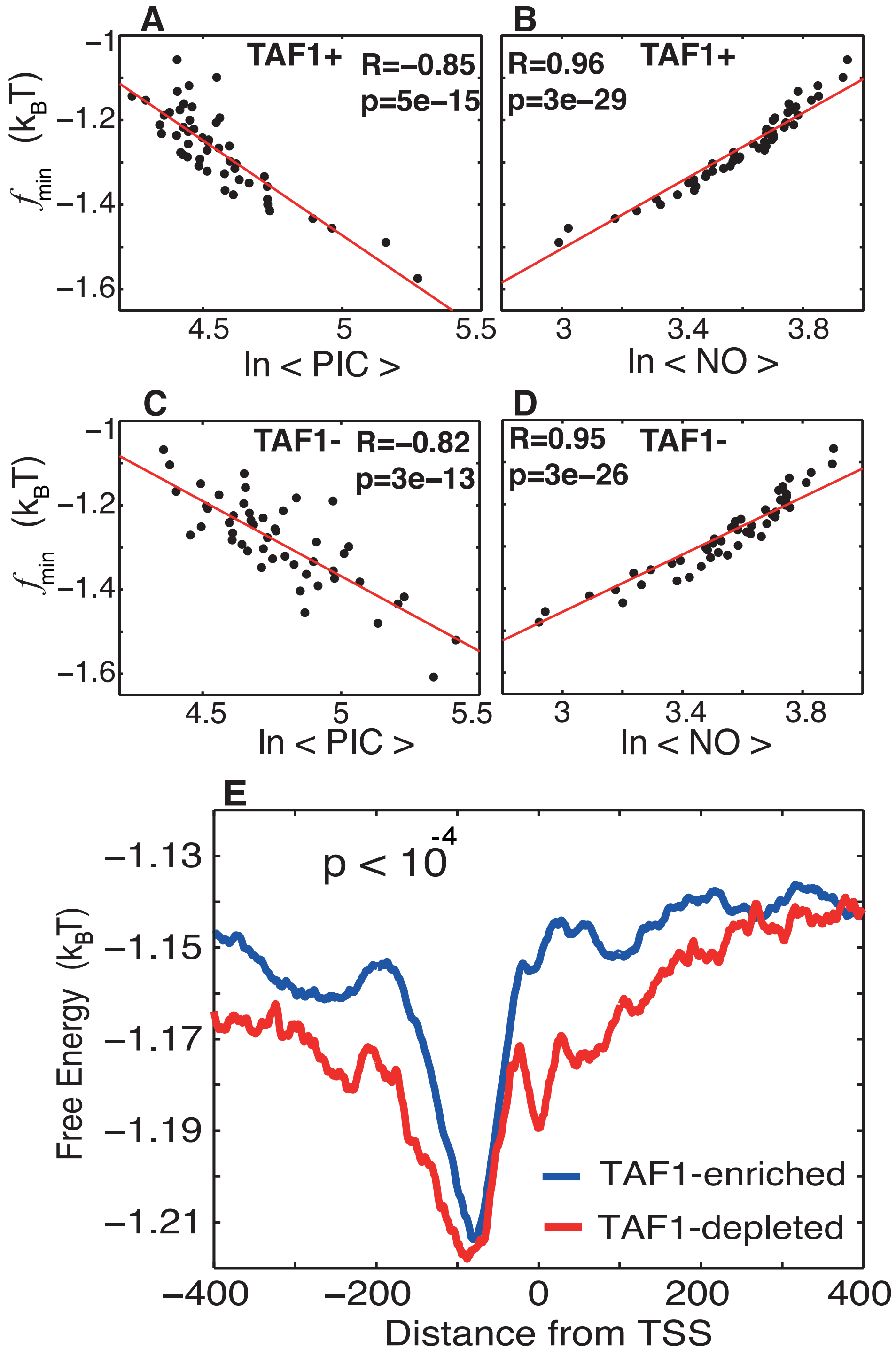

**Figure 5**

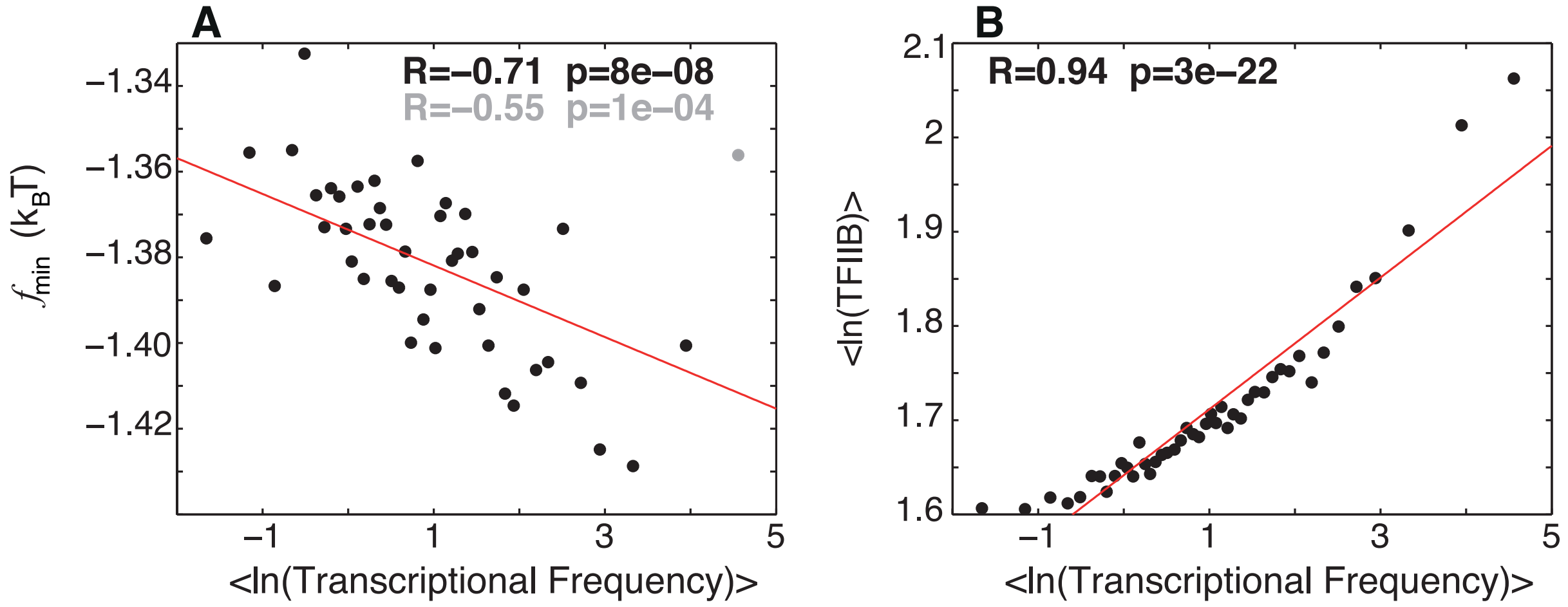

**Figure 6**

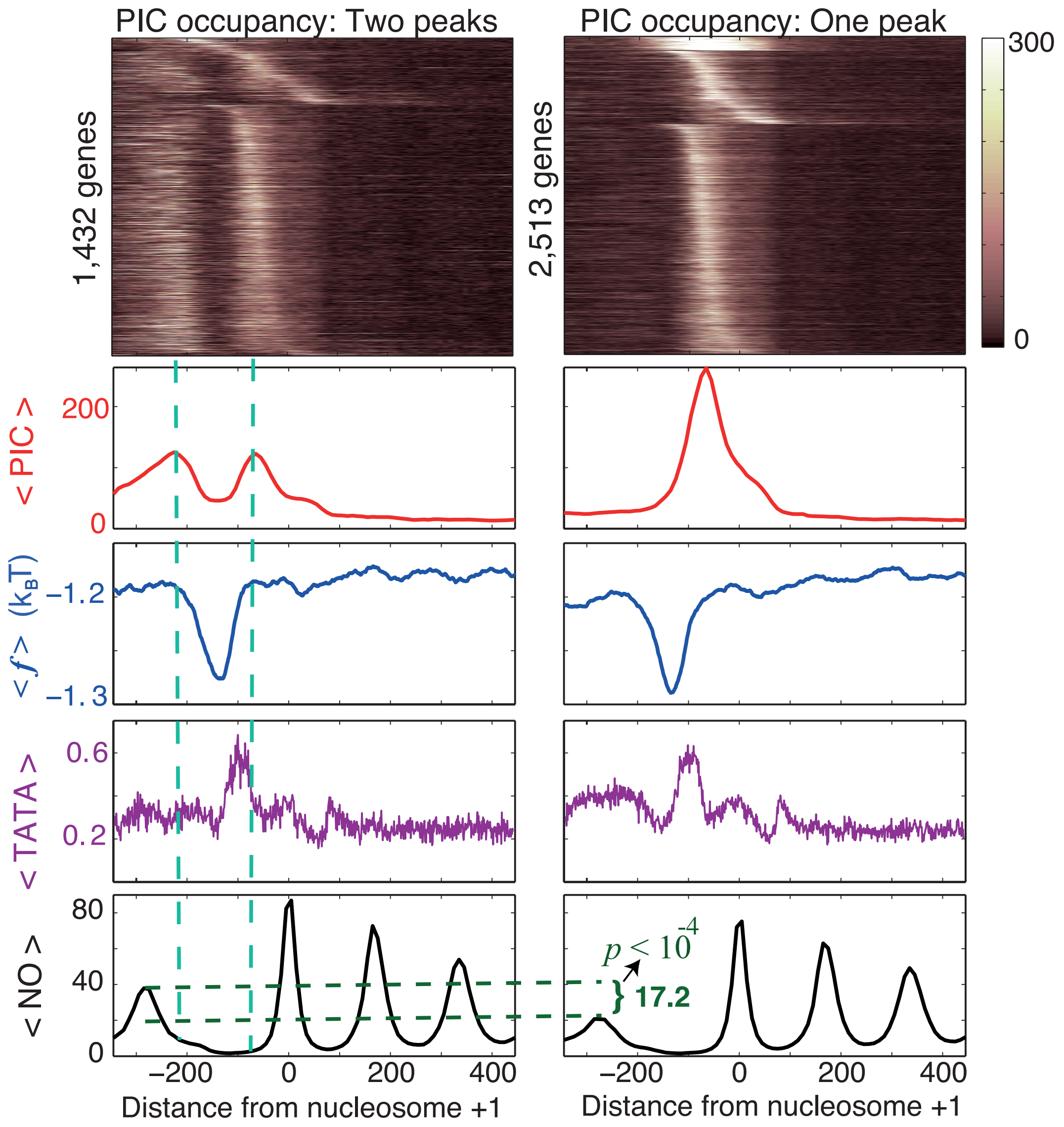

**Figure 7**

**Supporting Material**

# Genome-wide organization of eukaryotic pre-initiation complex is influenced by nonconsensus protein-DNA binding

Ariel Afek and David B. Lukatsky*
*Department of Chemistry, Ben-Gurion University of the Negev, Beer-Sheva 84015, Israel*

**Corresponding author:*
Email: lukatsky@bgu.ac.il

## Supporting Figure Legends

**Figure S1.** This figure demonstrates the robustness of the computed free energy of nonconsensus TF-DNA binding with respect to the global variability of the nucleotide content along the yeast genome (**A**); the robustness with respect to the width of the sliding window, $L$ (**B**); and the robustness with respect to the number of contacts, $M$, that the TF makes with DNA (**C**). (**A**) The average free energy of nonconsensus TF-DNA binding per bp, $\langle f \rangle = \langle \langle F \rangle_{\mathrm{TF}} \rangle_{seq} / M$ , (red), as compared with the corresponding normalized free energy per bp, $\langle \delta f \rangle = \langle \langle \delta F \rangle_{\mathrm{TF}} \rangle_{seq} / M$ , (blue), where $\delta F = F - F_{rand}$. For a given TF, $F$ is computed as described in the main text, and $F_{rand}$ is the free energy computed for a randomized sequence (in the same sliding window as $F$ ), and averaged over 25 random realizations. The described procedure removes the bias in the free energy, stemming from the global variability of the nucleotide content. (**B**) The normalized, average free energy, $\langle \delta f \rangle$, computed using different values of the width of the sliding window, $L = 30$ (red), $L = 50$ (black), and $L = 80$ (blue). We used $L = 50$ for all the calculations described in the main text. (**C**) The normalized, average free energy, $\langle \delta f \rangle$, computed using different values of the TF size, $M = 6$ (red), $M = 8$ (black), and $M = 10$ (blue). We used $M = 8$ for all the calculations described in the main text. In all plots, (**A**), (**B**), and (**C**), the average free energies are computed using 6,045 yeast transcripts.

**Figure S2.** This figure further demonstrates the statistical significance of the correlation between the free energy of nonconsensus TF-DNA binding and the PIC occupancy in the vicinity of the TSS. (**A**) Correlation between the minimal value of the free energy of nonconsensus TF-DNA binding, $f_{\min} = \min(f)$, where $f = \langle F \rangle_{\mathrm{TF}} / M$ , and the combined occupancy of all GTFs, computed for individual genes in non-overlapping windows of 80 bp within the entire interval $(-400, 400)$ around the TSS for each of these 3,945 genes. The data are binned into 50 bins. The notation $\langle \mathrm{PIC} \rangle$ describes the average, combined occupancy profile of all nine GTFs. (**B**) Correlation between the minimal value of the free energy of nonconsensus TF-DNA binding, $f_{\min} = \min(f)$, and the maximal combined occupancy of eight GTFs (all GTFs less the Pol II occupancy), computed for individual genes within the entire interval $(-150, 0)$ around the TSS for each of these 3,945 genes. The data are binned into 25 bins.

**Figure S3.** (**A**) Correlation between the minimal value of the free energy of nonconsensus TF-DNA binding, $f_{\min} = \min(f)$, where $f = \langle F \rangle_{\mathrm{TF}} / M$ , and the average, combined GTF occupancy computed for individual genes in non-overlapping windows of 80 bp within the entire interval $(-2990, 2070)$ around the open reading frame (ORF) ends for 2,903 mRNA genes. The data are binned into 50 bins. (**B**) Similarly computed correlation of $f_{\min}$ with the nucleosome occupancy. (**C**) The heat maps represent the free energy of nonconsensus TF-DNA binding, $f$, the combined occupancy of the GTFs, and the nucleosome occupancy, respectively, for 1,860 tandem mRNA genes aligned with respect to the ORF ends. The genes are sorted by intergenic length.

**Figure S4.** Correlation between the minimal value of the free energy of nonconsensus TF-DNA binding, $f_{\min} = \min(f)$, where $f = \langle F \rangle_{\mathrm{TF}} / M$ , and the average GTF occupancy

computed for individual genes, in non-overlapping windows of 80 bp within the entire interval $(-2990, 2070)$ around the ORF ends for 2,903 mRNA genes. The data are binned into 50 bins.

**Figure S5.** (**A**) Correlation between the minimal value of the free energy of nonconsensus TF-DNA binding, $f_{\min} = \min(f)$, where $f = \langle F \rangle_{\mathrm{TF}} / M$, and the average GTF occupancy of 676 TATA-containing genes. The correlation is computed for individual genes in non-overlapping windows of 80 bp within the entire interval $(-990, 990)$ around the TSS. The data are then binned into 50 bins. (**B**) Similar to (**A**), but now $f_{\min}$ is correlated with the nucleosome occupancy of these TATA-containing genes. (**C**) Correlation between $f_{\min}$, and the average GTF occupancy of 3,269 TATA-less genes. The correlation is computed for individual genes in non-overlapping windows of 80 bp within the entire interval $(-990, 990)$ around the TSS. The data are binned into 50 bins. (**D**) Similar to (**C**), but now $f_{\min}$ is correlated with the nucleosome occupancy of these TATA-less genes. (**E**) The average free energy of nonconsensus TF-DNA binding per bp, $\langle f \rangle$, for a larger set of 5,034 TATA-less genes (blue), and 1,011 TATA-containing genes (red), around the TSSs.

**Figure S6.** This figure is supplementary to **Figure 7** of the main text. It demonstrates the robustness of our conclusions with respect to an alternative definition of the TATA-like box occupancy score. (**A**) The TATA-like box occupancy scores were computed based on the PWM taken from V.X. Jin et al., BMC Bioinformatics **7**:114 (2006). We used 1,432 genes with double-peak in the nucleosome occupancy (red) and 2,513 genes with single-peak in the nucleosome occupancy (blue), as described in **Figure 7** of the main text. The obtained TATA-like box occupancy profiles are similar to those presented in **Figure 7** of the main text, and these profiles are statistically indistinguishable between those two groups of genes. (**B**) We also computed the initiator (INR) element PWM occupancy score for the selected two groups of genes, as described in (**A**). The obtained occupancy profiles are statistically indistinguishable between these two groups of genes.

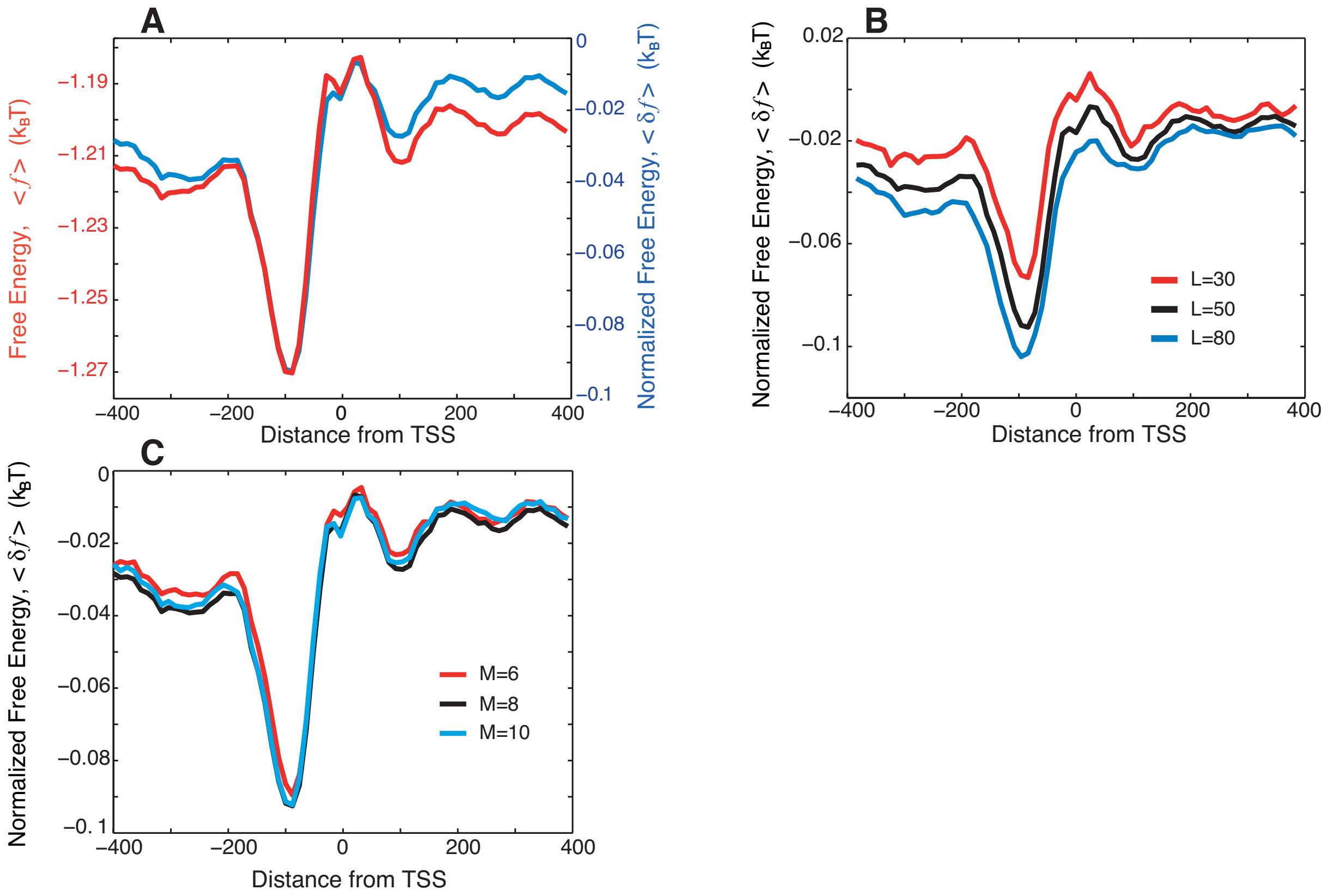

**Figure S1**

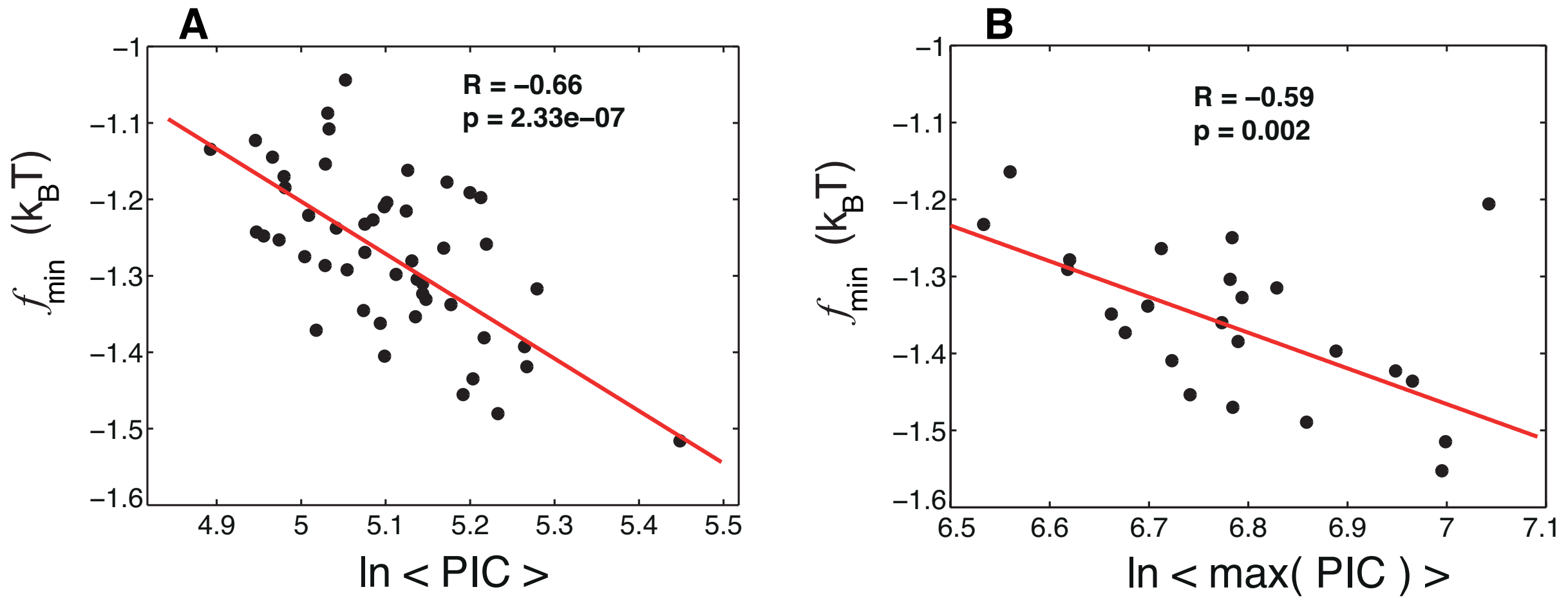

**Figure S2**

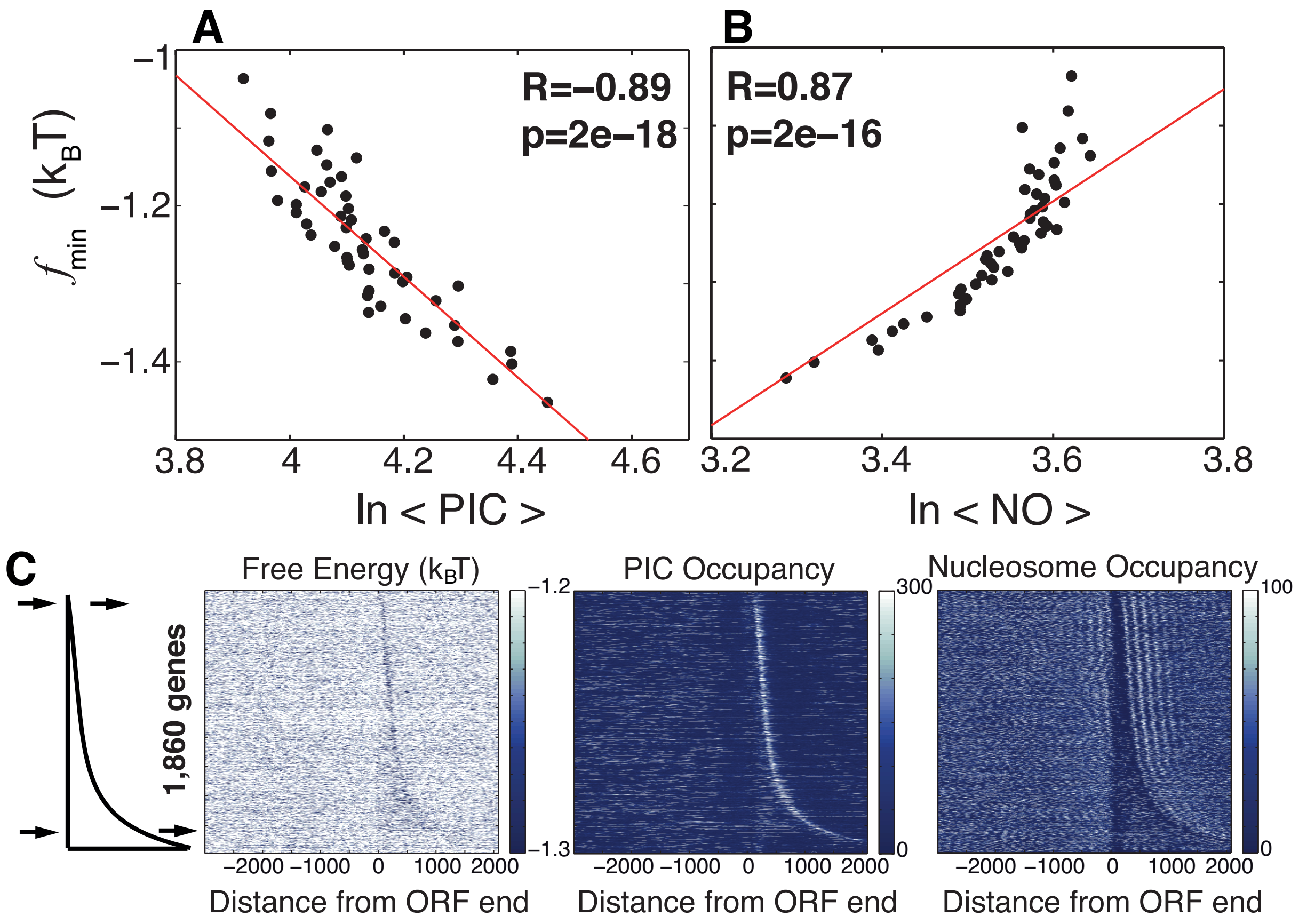
A
B
R=−0.89
p=2e−18
R=0.87
p=2e−16
$f_{min}$ ($k_BT$)
−1
−1.2
−1.4
3.8
4
4.2
4.4
4.6
3.2
3.4
3.6
3.8
ln < PIC >
ln < NO >
C
1,860 genes
Free Energy ($k_BT$)
−1.2
−1.3
PIC Occupancy
300
0
Nucleosome Occupancy
100
0
−2000 −1000 0 1000 2000
Distance from ORF end
−2000 −1000 0 1000 2000
Distance from ORF end
−2000 −1000 0 1000 2000
Distance from ORF end

**Figure S3**

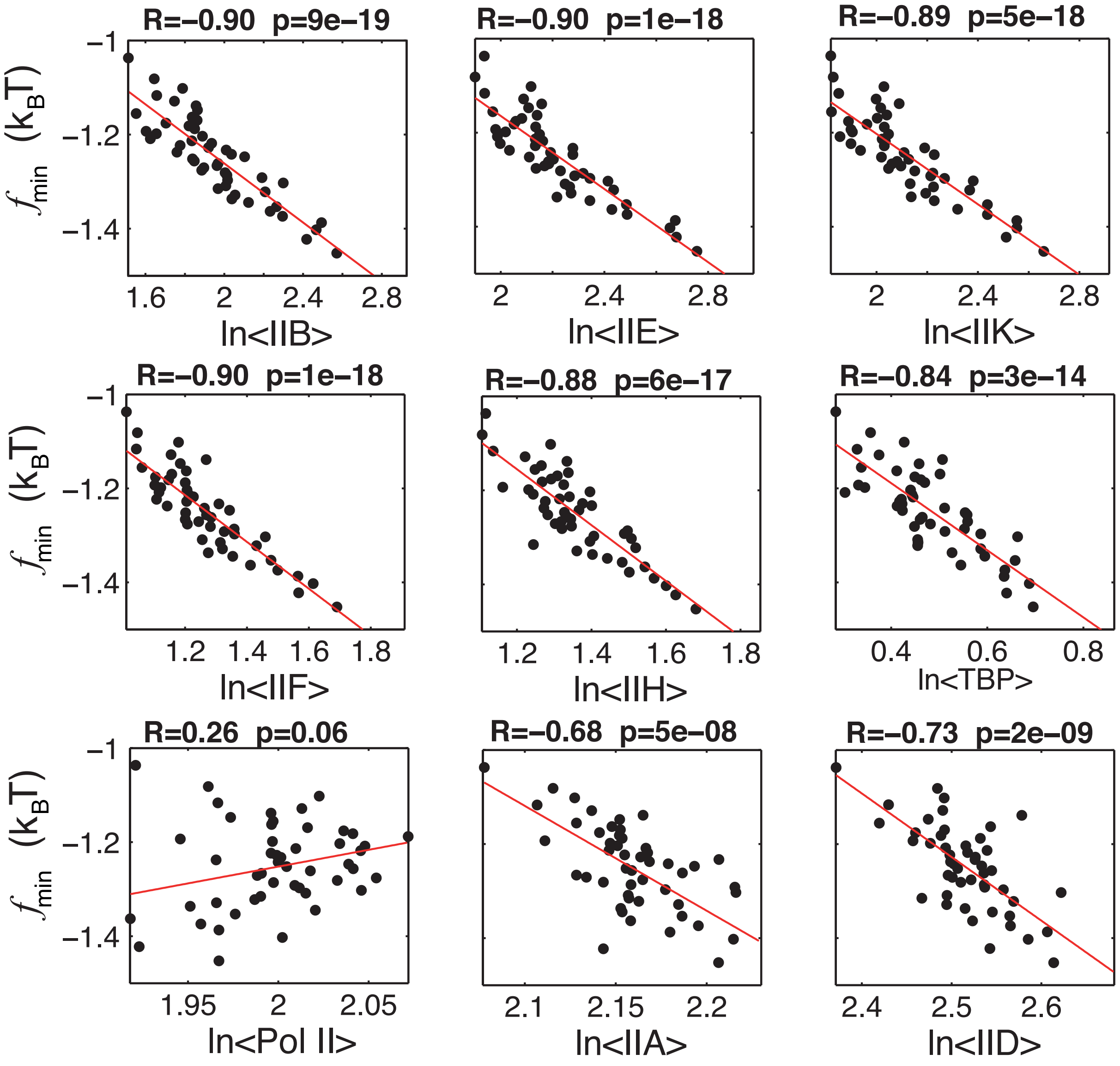

**Figure S4**

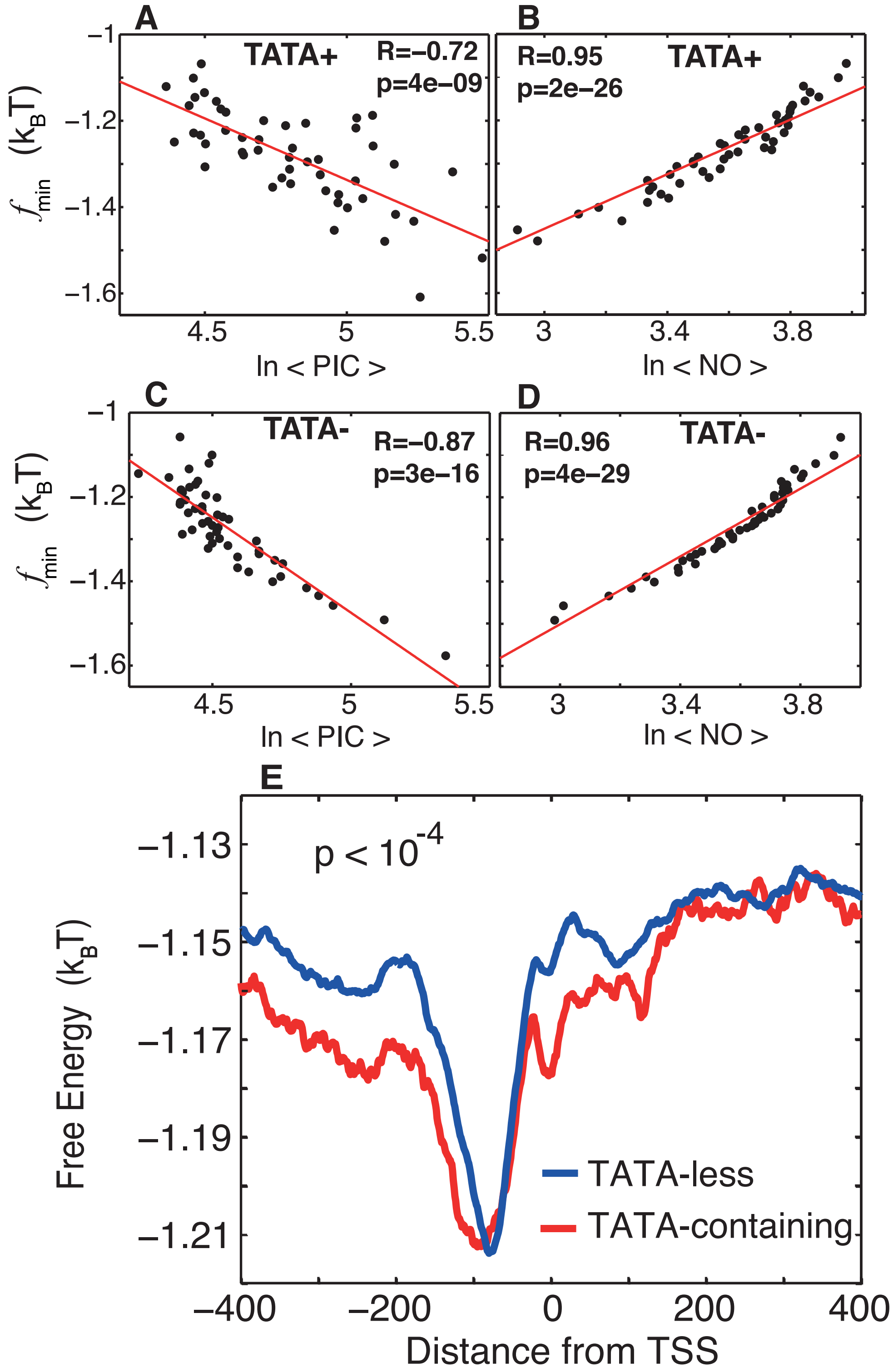

**Figure S5**

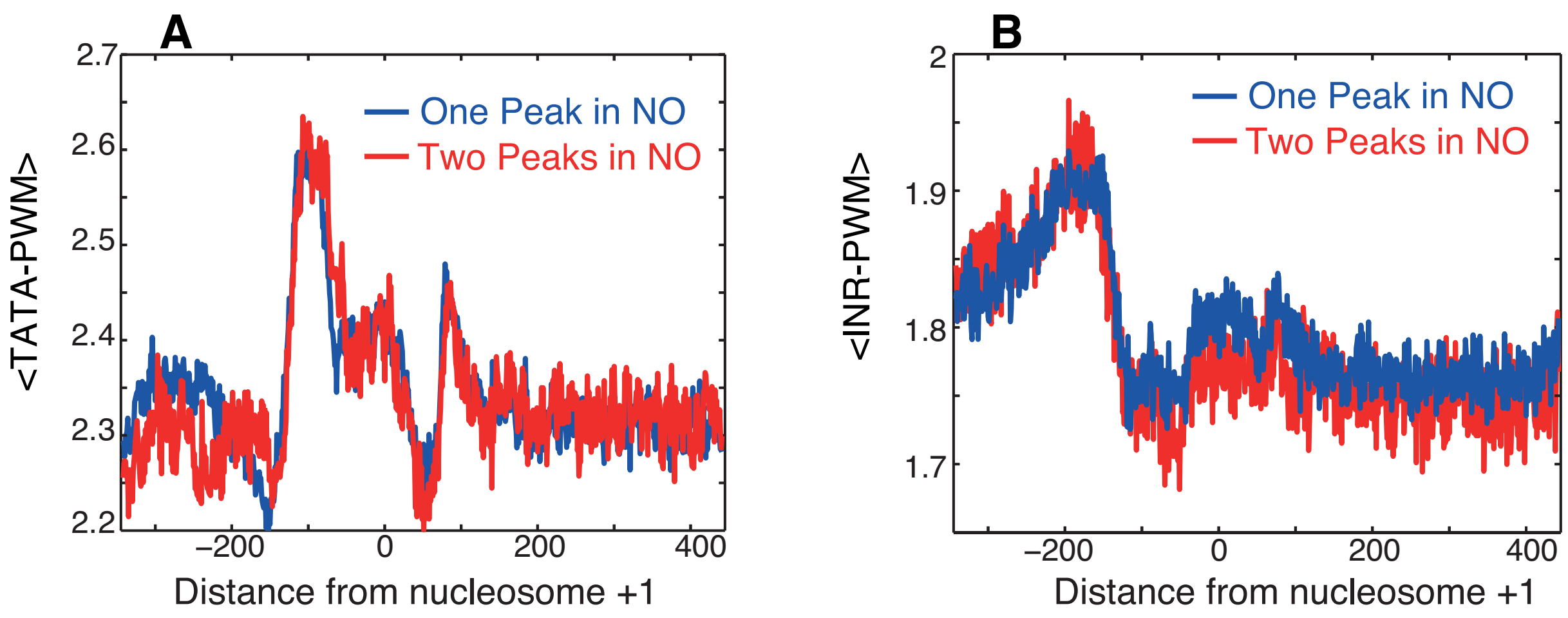

**Figure S6**